# Longitudinal Double Spin Asymmetry in Inclusive Jet Production at STAR

Katarzyna Kowalik for the STAR Collaboration

*Lawrence Berkeley National Laboratory, Berkeley, California 94720*

**Abstract.** This contribution reports on the first measurement of the longitudinal double-spin asymmetry $A_{LL}$ for the inclusive production of jets in polarized proton-proton collisions at $\sqrt{s} = 200\,\text{GeV}$. The data were collected with STAR at RHIC in the years 2003 and 2004, and correspond to a sampled integrated luminosity of $0.3\,\text{pb}^{-1}$ with beam polarizations up to 45%. The results on $A_{LL}$ cover jet transverse momenta $5 < p_T < 17\,\text{GeV/c}$ and agree with perturbative QCD evaluations based on deep-inelastic scattering parametrizations for the gluon polarization in the proton. The results disfavor large positive gluon polarization in the polarized proton.

## INTRODUCTION

The European Muon Collaboration (EMC) found that a surprisingly small fraction of the proton spin is carried by quark spins [1]. Subsequent polarized lepton-nucleon deep-inelastic scattering (DIS) experiments have confirmed the EMC data, and performed measurements with polarized deuteron and $^3$He targets and measurements covering complementary kinematic regions [2]. The polarization of gluons and the role of quark and gluon orbital momenta remain open questions to date.

Inclusive DIS is sensitive to gluon polarization, $\Delta G$, only through the dependence of the inclusive spin structure function on the virtuality of the photon, which is a higher order QCD effect. Due to the limited kinematic coverage of the polarized DIS experiments, the precision of $\Delta G$ from such an analysis is rather limited.

One way to improve the current knowledge of $\Delta G$ is via polarized semi-inclusive DIS measurements in which two high-$p_T$ hadrons or a charmed meson are detected [3]. Such processes are sensitive to gluons through photon-gluon fusion.

Another possibility is via polarized proton-proton collisions, where quark-gluon and gluon-gluon scattering provide the sensitivity. The Relativistic Heavy Ion Collider (RHIC) is the first polarized proton-proton collider, providing collisions at a center-of-mass energy $\sqrt{s} = 200\,\text{GeV}$ and in future $\sqrt{s} = 500\,\text{GeV}$ [4].

The STAR (Solenoidal Tracker at RHIC) experiment aims to determine the gluon polarization through several inclusive measurements and eventually via coincidence measurements of prompt photons and jets, $\vec{p} + \vec{p} \to \gamma + \text{jet}$. The initial focus is on inclusive processes with large production cross sections. At mid-rapidity and $\sqrt{s} = 200\,\text{GeV}$ these processes typically give sensitivity to the integral of $\Delta G$ over the range $0.03 \lesssim x \lesssim 0.3$ in the gluon momentum fraction. Coincidence measurements can give insight in the $x$-dependence of $\Delta G$, but require higher integrated luminosities than those that have been sampled to date. The STAR detector [5] features a large acceptance and is well-equipped for jet reconstruction at mid-rapidity.

These proceedings report on the first measurements of the longitudinal double-spin asymmetry $A_{LL}$ for inclusive jet production,

$$A_{LL} = \frac{\sigma^{++} - \sigma^{+-}}{\sigma^{++} + \sigma^{+-}}, \qquad (1)$$

where $\sigma^{++}$ ($\sigma^{+-}$) is the inclusive jet cross section when the colliding proton beams have equal (opposite) helicities.

The jet cross sections are described in a factorized form by a convolution of non-perturbative parton densities and hard-scattering matrix elements. Fragmentation functions do not enter the theoretical description. At jet transverse momenta $p_T \lesssim 8\,\text{GeV}/c$, the gluon-gluon scattering contribution to the polarized cross section is the largest. At higher $p_T$ in the range of the present measurements the quark-gluon scattering contribution is larger than the gluon-gluon contribution. Eventually, the quark-quark contribution becomes the dominant subprocess [6].

## EXPERIMENT AND ANALYSIS

During the initial proton running periods in the years 2003 and 2004, RHIC accelerated and stored typically 55 beam bunches, each with $5$–$7 \times 10^{10}$ protons, in each ring per store. Typical luminosities were $2$–$5 \times 10^{30}\,\text{cm}^{-2}\,\text{s}^{-1}$ and $\sqrt{s} = 200\,\text{GeV}$. Beam polarizations of up to 45% were reached. The beam polarizations were measured for each beam and for each beam fill with Coulomb-Nuclear Interference (CNI) proton-carbon polarimeters [7] which were calibrated *in situ* using a polarized atomic hydrogen gas-jet target [8].

The proton beam helicities at the STAR interaction region (IR) alternated for successive bunches in one ring and for successive pairs of bunches in the other ring. The beam bunches were separated by $\sim 213\,\text{ns}$. Hence, STAR recorded data with all four helicity combinations concurrently.

The STAR detector subsystems of interest here are the Time Projection Chamber (TPC), the Barrel Electromagnetic Calorimeter (BEMC), and the Beam-Beam Counters (BBC). The TPC is used to determine the momentum of charged particles in the pseudorapidity range $-1.3 \lesssim \eta \lesssim 1.3$ for all azimuthal angles $\phi$. The BEMC is a lead-scintillator sampling calorimeter which measured electromagnetic energy deposits for $0 < \eta < 1$ and $0 < \phi < 2\pi$. The BEMC installation was completed after the 2004 running period, and the detector presently covers $-1 < \eta < 1$. The BBCs are segmented scintillator annuli mounted around the beam pipe and span $3.3 < |\eta| < 5.0$. Coincident energy deposits in one or more segments of the BBC on either side of the STAR IR signal the collision of polarized protons at the IR and were used to form the so-called minimum bias (MB) trigger. In addition, the BBCs were used to measure proton beam luminosities and to measure transverse beam polarization components [9].

Jet data were collected with a high tower trigger (HT) that required an energy deposit in a BEMC tower ($\Delta \eta \times \Delta \phi = 0.05 \times 0.05$) in addition to the MB trigger condition. The trigger threshold corresponded to a transverse energy $E_T > 2.2\,\text{GeV}$ in 2003 and to $E_T > 2.2$–$3.4\,\text{GeV}$ for $\eta = 0 - 1$ in 2004. The integrated luminosity $\int \mathscr{L}\,dt$ amounts to

0.18 (0.12)pb$^{-1}$ for the analyzed 2003 (2004) data. A total of $3.0\times10^6$ HT events were analyzed.

Jets were reconstructed using a midpoint-cone algorithm [10] which clustered TPC tracks and BEMC energy deposits within a cone in $\eta$ and $\phi$ of radius of $R = 0.4$. The algorithm started from reconstructed TPC tracks and BEMC energy deposits above a seed cut of $0.2\,\text{GeV}$. Jets were merged if more than 50% of their energy was of common origin. A minimum jet $p_T > 5\,\text{GeV/c}$ was required.

Other selections in the analysis include the requirement of a reconstructed vertex on the beam axis within $\pm 60\,\text{cm}$ from the TPC center to ensure a uniform tracking efficiency. The jet axis was required to be within a fiducial range $0.2 < \eta < 0.8$ which, combined with the cone radius of $R = 0.4$, reduced the effects from BEMC acceptance edges. Beam background occasionally caused signal in the BEMC and its contribution was suppressed by requiring the ratio of jet energy in the BEMC to the total jet energy to be smaller than 0.8 (0.9) for the 2003 (2004) data sample.

The longitudinal double-spin asymmetry was evaluated according to:

$$A_{LL} = \frac{\sum(P_1 P_2)(N^{++} - R N^{+-})}{\sum(P_1 P_2)^2(N^{++} + R N^{+-})}, \qquad (2)$$

where $N^{++}$ ($N^{+-}$) is the jet yield for equal (opposite) proton beam helicities, $P_1 P_2$ is the product of the measured beam polarizations, and $R$ is the ratio of measured luminosities for equal and opposite proton beam helicities. The sums were performed over runs of about 20 minutes duration.

## RESULTS

The asymmetry $A_{LL}$ was evaluated separately for the HT jet sample of about $1\times 10^5$ jets collected in 2003, and the $2\times 10^5$ HT jets from 2004. The results on $A_{LL}$ from 2003 and 2004 data are in good agreement ($\chi^2/\text{ndf} = 0.3$) and cover $5 < p_T < 17\,\text{GeV/c}$.

Figure 1 shows the combined jet $A_{LL}$ versus jet $p_T$ with statistical error bars and two systematic uncertainty bands. The grey band shows the size of the combined scale uncertainty of 25% resulting from uncertainty in the beam polarization measurement. The hatched band shows the size of the total systematic uncertainty, which amounts to $\sim 0.01$. Sizable contributions to the systematic uncertainty arise from uncertainty in the measurement of $R$, from bias introduced by the HT trigger requirement and jet reconstruction, and from a possible contribution to the asymmetry measurement from residual non-longitudinal beam polarization components at the STAR IR through an azimuthally uniform transverse double-spin asymmetry [11]. Beam background forms a smaller contribution. Consistency checks with randomized beam-spin patterns and other checks including the evaluation of the parity-violating longitudinal single-spin asymmetries showed no evidence for beam bunch-to-bunch or fill-to-fill systematics in $A_{LL}$. A more detailed description may be found in Ref. [12].

The curves in Figure 1 show next-to-leading order perturbative-QCD (NLO pQCD) evaluations for $A_{LL}$ based on different polarized parton distribution functions [6]. The curve labeled GRSV-std is based on a best fit to inclusive DIS data [13]. The other curves use the assumption that the gluon polarization is maximal, $\Delta g = g$, that it is minimal, $\Delta g = -g$, or that it vanishes, $\Delta g = 0$, at the low initial scale of $0.4\,\text{GeV}^2/c^2$ of the parton parametrizations in Ref. [13]. The unpolarized cross section, discussed in a separate contribution [14], is reasonably well described by NLO pQCD over the measured range $5 < p_T < 50\,\text{GeV}/c$. This supports the use of NLO pQCD to constrain $\Delta G$ through measurement of $A_{LL}$. The STAR jet data on $A_{LL}$ are consistent with the DIS-based evaluation and disfavor saturated gluon polarization $\Delta g = g$ in the polarized proton ($\chi^2 \simeq 3$).

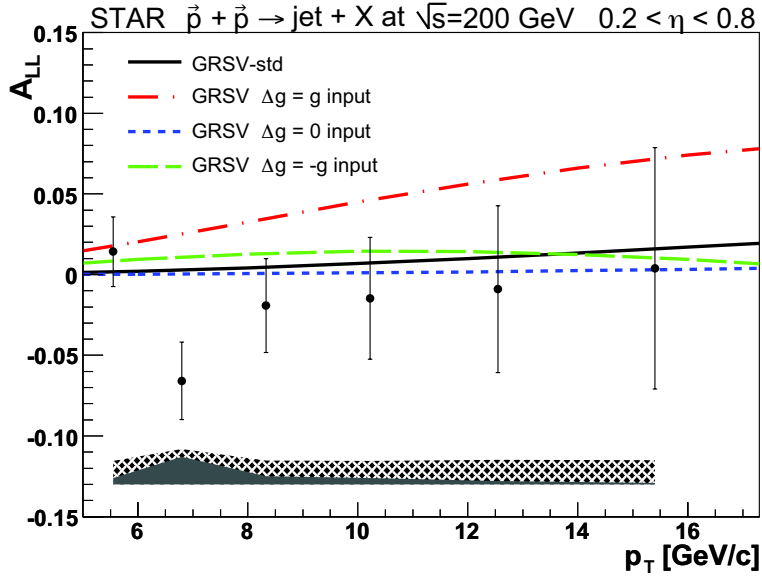

**FIGURE 1.** The longitudinal double-spin asymmetry $A_{LL}$ for inclusive jet production at $\sqrt{s} = 200\,\text{GeV}$ versus jet transverse momentum $p_T$. The error bars indicate the size of the statistical uncertainties. The inner band shows the size of the systematic uncertainty from the beam polarization measurement and the outer band shows the size of the total systematic uncertainty. The curves show different theoretical predictions for gluon polarization in the polarized proton [6] and are discussed in the text.

## SUMMARY AND OUTLOOK

We report the first measurement of the longitudinal double-spin asymmetry $A_{LL}$ for inclusive jet production in polarized proton collisions. The data cover jet transverse momenta $5 < p_T < 17\,\text{GeV}/c$ and were collected at $\sqrt{s} = 200\,\text{GeV}$ in the years 2003 and 2004. The uncertainties are statistics limited. The results are compared with pQCD calculations at NLO and disfavor large positive gluon polarization in the polarized proton.

In 2005, STAR has sampled an integrated luminosity of $3\,\text{pb}^{-1}$ with beam polarizations of about 50%, corresponding to an increase in figure-of-merit by an order of magnitude. The analysis is in progress and we anticipate results that will extend to higher $p_T$ and will distinguish extreme scenarios for the gluon polarization in the polarized nucleon.

The installation of the BEMC calorimeter was completed before the 2006 running period. The 2006 data was collected with the nominal STAR acceptance and should allow us to make polarized di-jet coincidence measurements, as well as to make a start on selective processes such as $\vec{p}+\vec{p} \to \gamma+\text{jet}$.

# REFERENCES

1. J. Ashman et al., *Nucl. Phys.* **B328**, 1 (1989).
2. SMC, B. Adeva et al., *Phys. Rev.* **D58**, 112001 (1998); E142, P.L. Anthony et al., *Phys. Rev.* **D54**, 6620 (1996); E143, K. Abe et al., *Phys. Rev.* **D58**, 112003 (1998); E154, K. Abe et al., *Phys. Rev. Lett.* **79**, 26 (1997); E155, P.L. Anthony et al., *Phys. Lett.* **B463**, 339 (1999); HERMES, K. Ackerstaff et al., *Phys. Lett.* **B404**, 383 (1997); HERMES, A. Airapetian et al., *Phys. Lett.* **B442**, 484 (1998); JLAB, X. Zheng et al., *Phys. Rev. Lett.* **92** 012004 (2004); COMPASS, E.S. Ageev et al., *Phys. Lett.* **B612**, 154 (2005).
3. HERMES, A. Airapetian et al., *Phys. Rev. Lett.* **84**, 2584 (2000); SMC, B. Adeva et al., *Phys. Rev.* **D70**, 012002 (2004); COMPASS, E. S. Ageev et al., *Phys. Lett.* **B633**, 25 (2006).
4. G. Bunce, N. Saito, J. Soffer and W. Vogelsang, *Ann. Rev. Nucl. Part. Sci.* **50**, 525 (2000).
5. Special Issue: RHIC and Its Detectors, *Nucl. Instrum. Meth.* **A499**, (2003).
6. B. Jäger, M. Stratmann, and W. Vogelsang, *Phys. Rev.* **D70**, 034010 (2004).
7. O. Jinnouchi et al., RHIC/CAD Note 171 (2004).
8. H. Okada et al., hep-ex/0601001, *Phys. Lett.* **B**, in press
9. J. Kiryluk et al., hep-ex/0501072, published the Spin 2004 conference proceedings, Trieste, Italy (2004).
10. G. C. Blazey et al., arXiv:hep-ex/0005012.
11. H. Mayer et al., *Phys. Rev. Lett.* **81**, 3096 (1998).
12. STAR, B.I. Abelev et al., hep-ex/0608030, *submitted to Phys. Rev. Lett.*
13. M. Glück, E. Reya, M. Stratmann and W. Vogelsang, *Phys. Rev.* **D63**, 094004 (2001)
14. M. Miller *for the STAR Collaboration*, these proceedings.